\documentclass{ws-ijmpcs}
\usepackage[super,compress]{cite}
\usepackage{graphicx}
\usepackage{bbm}
\usepackage{ulem}
\usepackage[colorlinks=true, pdfstartview=FitV,bookmarks=true, bookmarksnumbered=true, breaklinks]{hyperref}

\begin{document}

\markboth{M. Tanaka, D. Fujii, and A. Iwanaka}{Pion Gravitational Form Factors in Holographic QCD}

\newcommand{\bl}[1]{\textcolor{blue}{#1}}
\newcommand{\mg}[1]{\textcolor{magenta}{#1}}
\newcommand{\rd}[1]{\textcolor{red}{#1}}

%
\catchline{}{}{}{}{}
%

\title{Pion Gravitational Form Factors in Holographic QCD}

\author{Mitsuru~Tanaka}

\address{Department of Physics, Nagoya University, Nagoya, 464-8602, Japan\\
tanaka@hken.phys.nagoya-u.ac.jp}

\author{Daisuke~Fujii}

\address{Advanced Science Research Center, Japan Atomic Energy Agency (JAEA), Tokai, 319-1195, Japan\\
Research Center for Nuclear Physics, Osaka University, Ibaraki 567-0048, Japan\\
daisuke@rcnp.osaka-u.ac.jp}

\author{Akihiro~Iwanaka}

\address{Research Center for Nuclear Physics, Osaka University, Ibaraki 567-0048, Japan\\
iwanaka@rcnp.osaka-u.ac.jp}
\maketitle

\begin{history}
\published{Day Month Year}
\end{history}

\begin{abstract}
We present the first calculation of the momentum-transfer dependence of the pion gravitational form factors (GFFs) within a top-down holographic QCD framework. These form factors encode essential information about the internal stress distribution of hadrons and may serve as a tool to explore the non-perturbative dynamics responsible for quark and gluon confinement in QCD. In particular, we evaluate the forward-limit value of the GFFs, known as the D-term, within the framework of the Sakai–Sugimoto model. Our analysis also reveals glueball dominance, wherein the pion’s gravitational interaction is mediated by an infinite tower of glueball excitations.
\end{abstract}

\section{Introduction}

Understanding how quarks and gluons are confined within hadrons remains one of the most fundamental challenges in quantum chromodynamics (QCD), intimately connected with non-perturbative phenomena such as color confinement and spontaneous chiral symmetry breaking. In recent years, the study of the energy-momentum tensor (EMT), which encodes the spatial distribution of internal forces, has been explored as a possible means of gaining new insights into these mechanisms and the mechanical structure of hadrons~\cite{Polyakov:2018zvc}. The matrix elements of the EMT are encoded in gravitational form factors (GFFs), which characterize pressure, shear forces, spin, and energy distributions, and are now accessible via deeply virtual Compton scattering.

Notably, the internal force distribution in nucleons has been extracted experimentally for both quark and gluon sectors~\cite{Burkert:2018bqq,Duran:2022xag}, revealing a strong anisotropy: repulsive pressure dominates near the center, while attractive, confining forces emerge at larger distances. Alongside these experimental efforts, theoretical studies based on lattice QCD~\cite{Hackett:2023rif,HadStruc:2024rix,Hackett:2023nkr} and QCD effective model~\cite{Fujii:2024rqd,Fujii:2025aip,Sugimoto:2025btn} have provided valuable insights into the GFFs of nucleons and pions. 
These developments open the way for deeper understanding of how QCD dynamics give rise to the mechanical structure of hadrons.

In this proceedings, we report on the momentum-transfer dependence of pion GFFs using a top-down holographic QCD framework---the Sakai-Sugimoto (S-S) model~\cite{Sakai:2004cn,Sakai:2005yt}---following the same methodology as presented in Ref.~\refcite{Fujii:2024rqd}. While bottom-up holographic approaches~\cite{Abidin:2008hn} have also been used to study GFFs, the top-down construction allows for a derivation grounded in string theory via D-brane dynamics, which has been extensively studied for hadron spectroscopy~\cite{Constable:1999gb,Brower:2000rp,Sakai:2004cn,Hata:2007mb,Imoto:2010ef,Liu:2017xzo,Fujii:2020jre} and dynamical properties~\cite{Sakai:2005yt,Fujii:2021tsw,Liu:2021ixf,Iwanaka:2022uje,Fujii:2022yqh,Fujii:2023ajs}. In particular, we examine the D-term, which reflects the internal stress distribution and, in the soft pion limit, takes the universal value $D(0) = -1$ as a consequence of chiral symmetry~\cite{Donoghue:1991qv,Polyakov:1999gs,Hudson:2017xug}. This result sheds light on the non-perturbative structure of Nambu-Goldstone bosons and provides theoretical guidance for ongoing and future experimental investigations.

\section{Definition of the matrix element for the energy momentum tensor in the holographic QCD}

In this section, we describe the procedure for deriving the four-dimensional EMT in the context of top-down holographic QCD~\cite{Sakai:2004cn,Sakai:2005yt}, along with its precise definition. While this approach was originally developed in Ref.~\refcite{Fujita:2022jus} for spin-$1/2$ nucleons, we extend it here to the case of spin-0 pions.

The matrix elements of the EMT for a hadron with spin $0$ are given by
\begin{align}
    \big<\pi^a(p_2)|\hat{T}^{\mu\nu}(x)|\pi^b(p_1)\big> = \delta^{ab}\left[2P^\mu P^\nu A(t) + \frac{1}{2}\left(\Delta^\mu\Delta^\nu - \eta^{\mu\nu}\Delta^2\right)D(t)\right]e^{ix\cdot\Delta}, \label{GFFsspin0}
\end{align}
where $a, b$ are isospin indices, and the kinematic variables are defined as $P^\mu = (p_1^\mu + p_2^\mu)/2$, $\Delta^\mu = p_2^\mu - p_1^\mu$, and $t = \Delta^2$. We adopt the mostly-plus metric convention $\eta^{\mu\nu} = \mathrm{diag}(-1,+1,+1,+1)$ and normalize the single-particle states via $\langle \vec{p}_2 | \vec{p}_1 \rangle = 2E(2\pi)^3\delta^{(3)}(\vec{p}_1 - \vec{p}_2)$ with $E = \sqrt{\vec{p}_1^2 + m^2}$.

In the holographic framework, these matrix elements are obtained by analyzing how bulk metric perturbations, sourced by matter fields, manifest at the boundary through their asymptotic behavior. We employ the S-S model, in which $N_f$ D8-branes are embedded as probes into a background geometry sourced by $N_c$ D4-branes compactified on a circle~\cite{Witten:1998zw}. This background arises from a double Wick rotation of the $\mathrm{AdS}_7$ black hole $\times S^4$ geometry in 11-dimensional supergravity:
\begin{align}
    ds^2 = \frac{r^2}{L^2} \left[ f(r)d\tau^2 - dx_0^2 + dx_1^2 + dx_2^2 + dx_3^2 + dx_{11}^2 \right] + \frac{L^2}{r^2} \frac{dr^2}{f(r)} + \frac{L^2}{4} d\Omega_4^2, \label{AdS7BH}
\end{align}
where $f(r) = 1 - R^6/r^6$, $R = L^2 M_{KK}/3$, $L^3 = 8\pi g_s N_c l_s^3$, $R_{11} = g_s l_s = \lambda / (2\pi N_c M_{KK})$, and $\lambda N_c = L^6 M_{KK} / (32\pi g_s l_s^5) = 216\pi^3\kappa$. Here, $M_{KK}$ is the compactification scale for $\tau$, $g_s$ and $l_s$ are the string coupling and string length, respectively, and $R_{11}$ is the compactification radius in the $x^{11}$ direction.

The expectation value of the boundary EMT, $\langle T_{\mu\nu} \rangle$, is extracted from the subleading behavior of the bulk metric fluctuations. In particular, one finds
\begin{align}
    \int_{x_{11},\tau}\delta g_{\mu\nu} \sim \frac{2\kappa^2_7 L^5}{6} \frac{\langle T_{\mu\nu} \rangle}{r^4} + \cdots, \label{boundaryEMT}
\end{align}
where $\kappa_7^2$ is the gravitational constant in 7 dimensions, related to the 11-dimensional one via $\kappa^2_{11} = \mathrm{Vol}(S^4)\kappa^2_7 = \frac{8\pi^2}{3} \left(\frac{L}{2}\right)^4 \kappa^2_7$. This result is obtained using holographic renormalization in M-theory~\cite{deHaro:2000vlm}, which provides a technically advantageous framework compared to the non-conformal 10-dimensional case~\cite{Kanitscheider:2008kd}.

The metric perturbations $\delta g_{MN}$ satisfy the linearized Einstein equations:
\begin{align}
    \bar{\mathcal{H}}_{MN} &= \mathcal{H}_{MN} - \frac{1}{2} g_{MN} \mathcal{H}^P_P = -2\kappa_7^2 \mathcal{T}_{MN}, \label{linearEq} 
\end{align}
with $\mathcal{H}_{MN} \equiv \nabla^2 \delta g_{MN} + \nabla_M \nabla_N \delta g^P_P - \nabla^P \left( \nabla_M \delta g_{NP} + \nabla_N \delta g_{MP} \right) - \frac{12}{L^2} \delta g_{MN}$,
where indices $M, N, P$ run over the seven directions $(x^\mu, \tau, r, x^{11})$, after integrating over the $S^4$. The source term $\mathcal{T}_{MN}$ corresponds to the bulk EMT induced by the pion field. We work in axial gauge, imposing $\delta g_{Mr} = 0$.

To evaluate $\mathcal{T}_{MN}$, we consider the D8-brane action embedded in the 11-dimensional background:
\begin{align}
    &S_{D8}=-\frac{C}{2\pi R_{11}g_{s}}\int d\tau dx_{11}d^4xdrd\Omega_4(\delta(\tau)+\delta(\tau-\pi/M_{KK})) \notag \\
    &\hspace{30mm}\times\frac{\sqrt{-G_{(11)}}}{\sqrt{G_{\tau\tau}}}\frac{1}{4}G^{MN}G^{PQ}{\rm tr}[F_{MP}F_{NQ}]+...,
\end{align}
with $C = (64\pi^6 l_s^5)^{-1}$. The EMT is then defined as $\mathcal{T}_{MN} = -\frac{2}{\sqrt{-G_{(11)}}} \frac{\delta S_{D8}}{\delta G^{MN}_{(11)}} \mathrm{Vol}(S^4)$, 
leading to the following expression in four-dimensional Fourier space:
\begin{align}
    &\mathcal{T}_{MN}(k^\mu,r) \equiv 2\pi R_{11} \int d\tau'\, \mathcal{T}_{MN}(\tau', k^\mu, r), 
\label{bulkEMT}
\end{align}
where the radial coordinate $r$ is related to the holographic coordinate $z$ via $z = \pm \sqrt{r^6 / R^6 - 1}$.

In this analysis, we focus on the pion as the only dynamical source of gravitational backreaction. The bulk gauge fields $A_M(x^\nu, z)$ are expanded as
\begin{align}
    &A_\mu(x^\nu,z) = 0, \quad A_z(x^\nu,z) = \pi(x^\mu) \phi_0(z), \quad \phi_0(z) = c_0(1 + z^2)^{-1}, \label{pion}
\end{align}
with normalization condition $c_0^{-2} = 2 M_{KK}^2 \kappa \int dz\, (1 + z^2)\phi_0(z)^2$~\cite{Sakai:2004cn}. Since the pion appears as an external state, we substitute it with a plane wave $\pi(x^\nu) \sim e^{ip_\nu x^\nu}$ carrying momentum $p_\nu$.

More precisely, the GFFs are extracted from the three-point function $\langle J_A^\alpha T^{\mu\nu} J_A^\beta \rangle$, where $J_A^\alpha$ denotes the axial-vector current and $T^{\mu\nu}$ the EMT. This correlator is computed via the Gubser-Klebanov-Polyakov-Witten (GKP-Witten) prescription~\cite{Gubser:1998bc,Witten:1998qj}, and the pion contribution is isolated by applying the completeness relation of the axial-vector meson spectrum, as outlined in Ref.~\refcite{Abidin:2008hn}.

\section{Solving the Einstein equations}

To apply the matrix element definition in Eq.~\eqref{boundaryEMT}, we must solve the linearized Einstein equations~\eqref{linearEq} to determine the metric perturbations $\delta g_{MN}$.

Since we are interested in the EMT of QCD at the boundary, where the conservation law $\Delta^\mu \langle T_{\mu\nu} \rangle = 0$ must hold, we require the metric fluctuations to be transverse near the boundary, i.e., $\Delta^\mu \delta g_{\mu\nu} \sim 0$. These fluctuations are decomposed as $\delta g_{ij} \sim \delta g^{\rm TT}_{ij} + \left( \delta_{ij} - \frac{\Delta_i \Delta_j}{\vec{\Delta}^2} \right) \frac{\delta g^\alpha_\alpha}{5}$, 
where $\delta g^{\rm TT}_{ij}$ denotes the transverse-traceless (TT) part with $i,j = 1,2,3$, and $\alpha = \mu, \tau, 11$.

In the axial gauge, the Einstein equations impose the traceless condition in six dimensions, $\delta g^\alpha_\alpha \sim 0$, in the asymptotic AdS region ($r \rightarrow \infty$) with coordinates $(x^\mu, \tau, x^{11})$. This motivates us to focus on TT modes, which correspond to glueball excitations.

Among the 14 TT modes classified in Ref.~\refcite{Brower:2000rp} (see also Ref.~\refcite{Constable:1999gb}), only the $\mathrm{T}_4$ and $\mathrm{S}_4$ modes couple to the boundary EMT. By substituting the pion field~\eqref{pion} into Eq.~\eqref{gOther}, one finds that the contribution from $\delta g^{\rm other}_{\mu\nu}$ falls off more rapidly as $r \rightarrow \infty$ and can be neglected.

We decompose the metric fluctuations as $\delta g_{\mu\nu} = \delta g^{\rm T(2)}_{\mu\nu} + \delta g^{\rm T(0)}_{\mu\nu} + \delta g^{\rm S(0)}_{\mu\nu} + \delta g^{\rm other}_{\mu\nu}$. 
The Einstein equations then reduce to
\begin{align}
    &\frac{r^2}{L^2} \left[ \frac{1}{L^2 r^5} \partial_r \left( (r^7 - r R^6) \partial_r \right) - \frac{L^2 \Delta^2}{r^2} \right] \frac{L^2}{r^2} \delta g^{\rm T}_{\mu\nu} = -2 \kappa_7^2 \mathcal{T}^{\rm T}_{\mu\nu}, \\
    &\frac{r^2}{L^2} \left[ \frac{1}{L^2 r^5} \partial_r \left( (r^7 - r R^6) \left( \partial_r + \frac{144 r^5 R^6}{(5 r^5 - 2 R^6)(r^6 + 2 R^6)} \right) \right) \right. \notag \\
    &\hspace{34mm} \left. - \frac{L^2 \Delta^2}{r^2} \right] \frac{L^2}{r^2} \delta g^{\rm S(0)}_{\mu\nu} \Big|_{r \rightarrow \infty} = -2 \kappa_7^2 \mathcal{T}^{\rm S(0)}_{\mu\nu},
\end{align}
where $\delta g^{\rm T}_{\mu\nu} \equiv \delta g^{\rm T(2)}_{\mu\nu} + \delta g^{\rm T(0)}_{\mu\nu}$, and the source term is decomposed as $\mathcal{T}_{\mu\nu} = \mathcal{T}^{\rm T}_{\mu\nu} + \mathcal{T}^{\rm S(0)}_{\mu\nu} + \mathcal{T}^{\rm other}_{\mu\nu}$.

Following Ref.~\refcite{Fujita:2022jus}, each component of the source term is given by
\begin{align}
    &2\kappa^2_7 \mathcal{T}^{\rm T}_{\mu\nu} = 2\kappa^2_7 \mathcal{T}_{\mu\nu} - 2\kappa^2_7 \mathcal{T}^{\rm S}_{\mu\nu} - 2\kappa^2_7 \mathcal{T}^{\rm other}_{\mu\nu}, \\
    &2\kappa^2_7 \mathcal{T}^{\rm S}_{\mu\nu} = \frac{r^6 + 2R^6}{4(r^6 - R^6)} \left( \eta_{\mu\nu} - \frac{\Delta_\mu \Delta_\nu}{\Delta^2} \right) \notag \\
    &\quad \times \left[ \frac{1}{L^4 r^3} \partial_r \left( r(r^6 - R^6)\partial_r(a + b) - 3R^6(a + b) \right) - \Delta^2 b \right], \\
    &2\kappa^2_7 \mathcal{T}^{\rm other}_{\mu\nu} = \frac{1}{L^4 r^3} \partial_r \left[ \left( \eta_{\mu\nu} - \frac{\Delta_\mu \Delta_\nu}{\Delta^2} \right) r(r^6 - R^6)\partial_r a - 3R^6 \eta_{\mu\nu} b \right],
\end{align}
with auxiliary functions $a(r, \Delta)$ and $b(r, \Delta)$ defined through $\partial_r a = -\frac{3R^6}{(5r^6 - 2R^6)} b + \frac{18r^{11}}{(5r^6 - 2R^6)R^6} \kappa^2_7 \mathcal{T}_{zz}$ and $b = -i \frac{\Delta^\mu}{\Delta^2} \frac{r^6 \sqrt{r^6 - R^6}}{R^9} 2\kappa^2_7 \mathcal{T}_{\mu z}$. 
The corresponding metric perturbations are parametrized as
\begin{align}
    &\delta g^{\rm other}_{\mu\nu}(\vec{\Delta}, r) = \frac{r^2}{L^2} \frac{\Delta_\mu \Delta_\nu}{\Delta^2} a(r, \Delta), \notag \\
    &\delta g^{\rm other}_{11,11}(\vec{\Delta}, r) = \frac{r^2}{L^2} b(r, \Delta), \quad \delta g^{\rm other}_{\tau\tau}(\vec{\Delta}, r) = \frac{r^2 f(r)}{L^2} \left( -b(r, \Delta) \right). \label{gOther}
\end{align}

The metric perturbations sourced by the pion field can be obtained using the Green’s function method:
\begin{align}
    \delta g^{\rm T/S}_{\mu\nu}(\vec{\Delta}, r) \sim \frac{r^2}{L^2} \int dr' \sqrt{-G_{(7)}(r')} G^{\rm T/S}(\vec{\Delta}, r, r') \frac{L^2}{{r'}^2} 2\kappa^2_7 \mathcal{T}^{\rm T/S}_{\mu\nu}(\vec{\Delta}, r'), \label{solg}
\end{align}
where $\sim$ denotes the asymptotic form in the $r \rightarrow \infty$ limit for the ${\rm S}_4$ mode. The Green's function is expressed as $G^{\rm T/S}(\vec{\Delta}, r, r') = \sum_{n=1}^\infty \frac{\Psi^{\rm T/S}_n(r) \Psi^{\rm T/S}_n(r')}{(m^{\rm T/S}_n)^2 + \vec{\Delta}^2}$, 
in terms of the eigenfunctions $\{\Psi_n^{\rm T/S}(r)\}$ of glueballs satisfying
\begin{align}
    &\frac{1}{r^3 L^4} \partial_r \left( (r^7 - r R^6) \partial_r \Psi^{\rm T}_n \right) = - (m^{\rm T}_n)^2 \Psi^{\rm T}_n, \\
    &\frac{1}{r^3 L^4} \partial_r \left( (r^7 - r R^6) \left( \partial_r + \frac{144 r^5 R^6}{(5r^6 - 2R^6)(r^6 + 2R^6)} \right) \Psi^{\rm S}_n \right) = - (m^{\rm S}_n)^2 \Psi^{\rm S}_n. \label{EVeq}
\end{align}
The eigenfunctions obey the completeness relations:
\begin{align}
    \frac{r^3}{L^3} \sum_{n=1}^\infty \Psi^{\rm T/S}_n(r) \Psi^{\rm T/S}_n(r') = \delta(r - r'), \quad
    \int_R^\infty dr \frac{r^3}{L^3} \Psi^{\rm T/S}_n(r) \Psi^{\rm T/S}_m(r) = \delta_{nm}. \label{CR}
\end{align}

Finally, we note that the dominant contribution of the EMT in the bulk propagated by each of the T(2), T(0) and S(0) states are $\mathcal{T}_{\mu\nu}$, $\mathcal{T}_{11,11}$ and $\mathcal{T}_{\tau\tau}$ respectively. 
Although Eq.~\eqref{bulkEMT} ensures that the total $\mathcal{T}_{\tau\tau}$ vanishes, the component $\mathcal{T}_{\tau\tau}^{\rm S}$, which satisfies the relation $\mathcal{T}_{\tau\tau}^{\rm S}+\mathcal{T}_{\tau\tau}^{\rm other}=0$, serves as the relevant source term for the ${\rm S}_4$ mode in the bulk.

\section{Results for the GFFs from top-down holographic QCD}

By substituting the pion field into the bulk EMT, we extract the matrix elements of the boundary EMT via Eq.~\eqref{boundaryEMT} and the holographic dictionary~\eqref{solg}, yielding
\begin{align}
    &A(t) = \frac{6}{L^7} \int^\infty_R dr^\prime \sum^\infty_{n=1} \frac{\alpha^{\rm T}_n \Psi^{\rm T}_n(r^\prime)}{(m^{\rm T}_n)^2 + \vec{\Delta}^2} \frac{{r^\prime}^3}{L^3} P^{\mu\nu}_{\rm t} \mathcal{T}_{\mu\nu}(\vec{\Delta}, r^\prime), \\
    &D(t) = -\frac{6}{L^7} \int^\infty_R dr^\prime \sum^\infty_{n=1} \frac{\alpha^{\rm T}_n \Psi^{\rm T}_n(r^\prime)}{(m^{\rm T}_n)^2 + \vec{\Delta}^2} \frac{{r^\prime}^3}{L^3} P^{\mu\nu}_{\rm s} \left( \mathcal{T}_{\mu\nu} + \mathcal{T}^{\rm S}_{\mu\nu} + \mathcal{T}^{\rm other}_{\mu\nu} \right) \notag \\
    &\hspace{12mm} + \frac{6}{L^7} \int^\infty_R dr^\prime \sum^\infty_{n=1} \frac{\alpha^{\rm S}_n \Psi^{\rm S}_n(r^\prime)}{(m^{\rm S}_n)^2 + \vec{\Delta}^2} \frac{{r^\prime}^3}{L^3} P^{\mu\nu}_{\rm s} \mathcal{T}^{\rm S}_{\mu\nu}, \\
\end{align}
where $P^{\mu\nu}_{\rm t} = (3P^\mu P^\nu / P^2 - \eta^{\mu\nu}) / P^2$ and $P^{\mu\nu}_{\rm s} = (P^\mu P^\nu / P^2 - \eta^{\mu\nu}) / \Delta^2$ are projection tensors onto traceless and scalar components, respectively, and $\alpha^{\rm T/S}_n$ is defined as $\alpha^{\rm T/S}_n = \frac{(m^{\rm T/S}_n)^2 L^4}{6} \int^\infty_R dr\, r^3 \Psi^{\rm T/S}_n(r)$.

To evaluate the infinite sums in the Green's function, we expand them as a power series in $-\vec{\Delta}^2$:
\begin{align}
    &\frac{6}{L^7}  \sum^\infty_{n=1} \frac{\alpha^{\rm T/S}_n \Psi^{\rm T/S}_n(r^\prime)}{(m^{\rm T/S}_n)^2 + \vec{\Delta}^2} = \sum^\infty_{k=0} F_k^{\rm T/S}(r^\prime)(-\vec{\Delta}^2)^k, 
\label{expand}
\end{align}
with $F^{\rm T/S}_k(r^\prime) = \int_R^\infty dr \frac{r^3}{L^3} \sum_{n=0}^\infty \frac{\Psi^{\rm T/S}_n(r) \Psi^{\rm T/S}_n(r^\prime)}{(m_n)^{2k}}$. The functions $F^{\rm T/S}_k(r)$ satisfy the following recurrence relations derived from the eigenvalue equations~\eqref{EVeq}:
\begin{align}
    &\frac{1}{r^3 L^4} \partial_r \left( (r^7 - r R^6) \partial_r F^{\rm T}_k \right) = -F^{\rm T}_{k-1}, \\
    &\frac{1}{r^3 L^4} \partial_r \left( (r^7 - r R^6) \left( \partial_r + \frac{144 r^5 R^6}{(5r^6 - 2R^6)(r^6 + 2R^6)} \right) F^{\rm S}_k \right) = -F^{\rm S}_{k-1},
\end{align}
with the initial condition $F^{\rm T/S}_0 = 1$ ensured by the completeness relation~\eqref{CR}. These functions are obtained recursively by imposing the boundary conditions $\partial_r F^{\rm T/S}_k(r)|_{r \to R} = 1$ and $F^{\rm T/S}_k(r)|_{r \to \infty} = 0$.

\begin{figure}
    \includegraphics[scale=0.2]{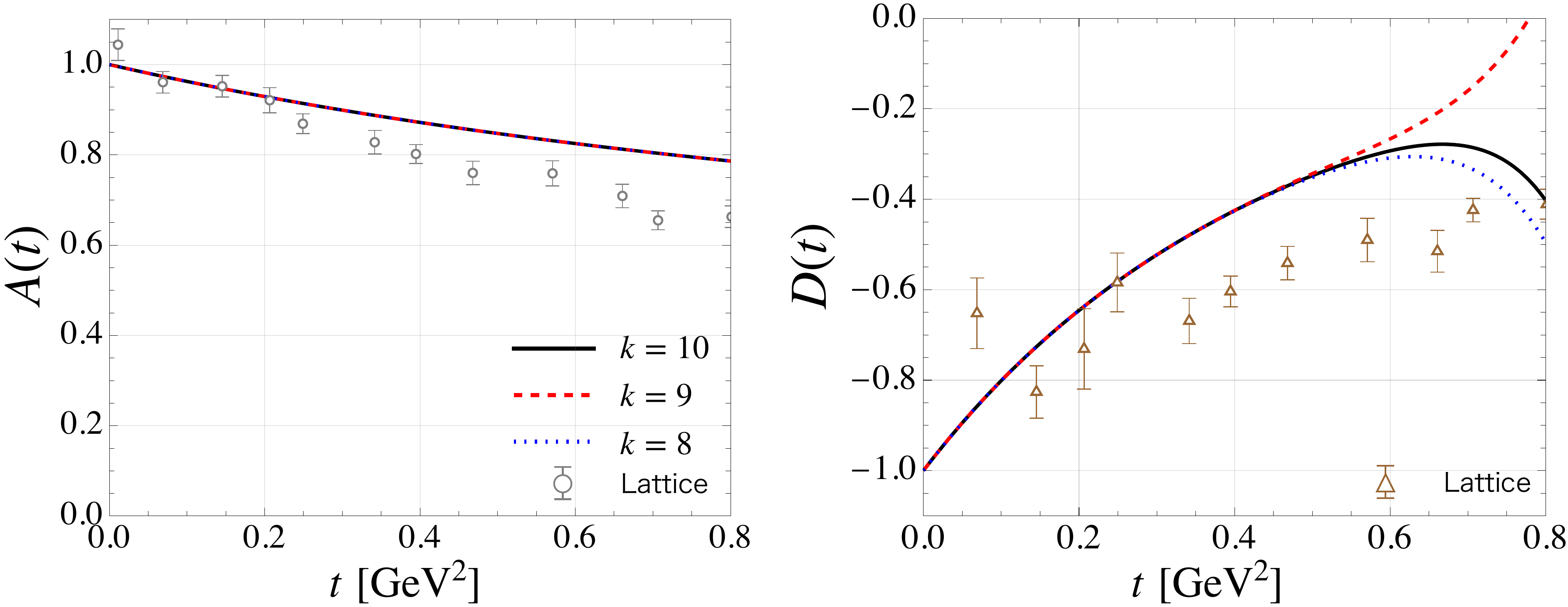}
    \caption{\label{AtDt} The pion GFFs $A(t)$ and $D(t)$ compared with lattice QCD results~\cite{Hackett:2023nkr}. For details, see Ref.~\protect\refcite{Fujii:2024rqd}}
\end{figure}

Figure~\ref{AtDt} shows the resulting GFFs $A(t)$ and $D(t)$ obtained by truncating the expansion~\eqref{expand} at $k=10$ (also showing results for $k=8,9$). We adopt the standard parameters in the S-S model $\kappa = 0.00745$ and $M_{KK} = 949\,\mathrm{MeV}$, determined from the $\rho$ meson mass $776\,\mathrm{MeV}$ and the pion decay constant $f_\pi = 92.4\,\mathrm{MeV}$~\cite{Sakai:2005yt}.

We observe that $A(t)$ exhibits good convergence for $t \lesssim 0.8\,\mathrm{GeV}^2$, while $D(t)$ begins to deviate from convergence at $t \gtrsim 0.5\,\mathrm{GeV}^2$. These deviations become significant as $t$ approaches the model's cutoff scale $M_{KK}^2$, beyond which the applicability of the holographic framework is limited and higher precision is not pursued.

Importantly, our top-down calculation reveals that $D(t)$ decreases more rapidly than $A(t)$, in qualitative agreement with recent lattice QCD results~\cite{Hackett:2023nkr}. This behavior arises from the additional contribution of the ${\rm S}_4$ glueball mode to $D(t)$, which is absent in bottom-up models such as Ref.~\refcite{Abidin:2008hn}, where there is no $\mathcal{T}_{\tau\tau}$ source coupling to scalar glueballs. 
Consistent with previous findings~\cite{Fujita:2022jus}, our results support the glueball dominance picture, where pion GFFs are governed by an infinite tower of glueball excitations in the bulk.

Finally, we obtain the forward-limit value $D(0) = -1$, as shown in Fig.~\ref{AtDt}. This result reflects loop-level contributions in the large-$N_c$ limit and contrasts with $A(0) = 1$, which follows from general constraints. The value $D(0) = -1$ is in accordance with the soft-pion theorem in the chiral limit~\cite{Donoghue:1991qv,Polyakov:1999gs,Hudson:2017xug}, as expected in our setup. Several proposals have been made to include quark mass effects in the S-S model~\cite{Casero:2007ae,Hashimoto:2007fa}, and their impact on the GFFs remains an interesting subject for future investigation.

\section{Summary and outlook}

In this work, we have calculated the GFFs of pions within a top-down holographic QCD framework. By solving the bulk Einstein equations with the pion field as the source of gravitational perturbations, and employing the holographic correspondence~\eqref{boundaryEMT}, we have extracted the matrix elements of the EMT. Our results exhibit qualitative agreement with recent lattice QCD findings~\cite{Hackett:2023nkr}, particularly reproducing the distinct momentum-transfer dependence of $A(t)$ and $D(t)$, a feature not observed in bottom-up models~\cite{Abidin:2008hn}. In the chiral limit, we also obtained the D-term as $D(0) = -1$.

This analysis suggests multiple directions for further investigation. One natural extension is to apply the same method to other hadrons, such as non-zero spin particles including nucleons and vector mesons. Another important avenue is to explore how these results are modified beyond the chiral limit, especially in light of previous discussions connecting the Gell-Mann––Oakes––Renner relation with mechanical stability conditions~\cite{Son:2014sna}. Furthermore, 
as demonstrated in Ref.~\refcite{Fujii:2025aip}, investigating the connection between the confining pressure and the trace anomaly may also offer further insight into the nonperturbative structure of QCD.

\section*{Acknowledgements}

We are grateful to the authors of Ref.~\refcite{Hackett:2023nkr} for generously supplying the data tables utilized in FIG.~\ref{AtDt}.
This work was supported by the Japan Society for the Promotion of Science (JSPS) KAKENHI (Grants No. JP24K17054, No. JP24KJ1620).

\bibliographystyle{ws-ijmpcs}
\bibliography{ref}










\end{document}